# MONOTONIC AND FATIGUE TESTING OF SPRING-BRIDGED FREESTANDING MICROBEAMS APPLICATION FOR MEMS

*Ming-Tzer Lin, Kai-Shiang Shiu, Chi-Jia Tong*

Institute of Precision Engineering, National Chung Hsing University, Taichung, 402, Taiwan

## ABSTRACT

An electroplating spring-bridge micro-tensile specimen is fabricated to carry out a series of monotonic tensile testing on it. Freestanding thin films were loaded by performing monotonic loading/unloading and tension-tension fatigue experiments. Loading was applied using a piezoelectric actuator with 0.1 μm resolution connected through pin hole into the test chip specimen. Loads were measured by connected a capacitor load cell with a resolution of less than 0.1 mN. We found the modulus of gold, copper and tantalum nitride thin films with thickness of 200~800nm at ambient temperature. Displacement controlled tension-tension fatigue experiments have also been performed and a trend of decreasing cycles to failure with increasing displacement amplitude and increasing mean displacement has been noted.

## 1. INTRODUCTION

Microelectromechanical systems (MEMS) technologies are developing rapidly with increasing study of the design, fabrication and commercialization of microscale systems and devices. Accurate knowledge on the mechanical behaviors of thin film materials used for MEMS is important for successful design and development of MEMS.

One of the MEMS application are primary used in RF switch, which has a potential use in tunable transmission lines for cellular telephone applications, or phase array radars. Advantages of RF switches are excellent linearity up to high switching speeds and low power consumption. In an RF switch, schematically depicted in Figure 1, an electrostatic force applied between the substrate and a bridge structure pulls the bridge down. This causes a change in capacitance by a factor of 100 or more. The state in which the electrostatic force is applied corresponds to a closed circuit (the "on" position) and the state with no electrostatic force applied corresponds to an open circuit, or the "off" position. During operation of an RF switch, the bridge structures deflected as far as the gap permits. Since these deflections are anticipated to reach kHz frequencies, very high cycle numbers can be attained over short time periods.

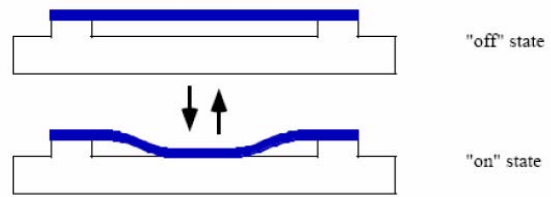

Fig. 1. The Schematic of RF switches operation.

Thus, understanding of fundamental observed failure mode and mechanical response respected to external loads plays an important role in products design and lifetime prediction and needs to be established.

Here an electroplating micro-tensile specimen integrates pin-pin align holes, misalignment compensate spring, load sensor beam and freestanding thin film is demonstrated and fabricated. The specimen is fit into a specially designed micro-mechanical apparatus to carry out a series of monotonic tensile testing as well as displacement controlled tension-tension fatigue experiments on sub-micron freestanding thin films is demonstrated.

## 2. EXPERIMENTAL PROCEDURES

Thin films applicable as structure or motion gears in MEMS were tested including sputtered gold, copper and tantalum nitride thin films. Metal specimens were fabricated by sputtering; for tantalum nitride film samples, nitrogen gas was introduced into the chamber during sputtering tantalum films on the silicon wafer.

A typical fabricated specimen ready for testing is shown in Fig.2. The force sensor beam was used to measuring the stress of freestanding thin films. The spring structure (including two supported beams and U shape spring), which have the ability to reduce the misalign errors by 6 orders of magnitude [6]. There are four principal pin holes in samples, bigger two in both ends were fixed to actuator and stage, and others were connected to capacity sensors.

FEM simulation was performed to verify the sample design concepts and preliminary calculation for load cell compliance as well as stress distribution for the specimens.

Fig. 3 shows the stress uniformly distributed on the bridge section of thin films while exerting the tensile loading using ANASYS simulation.

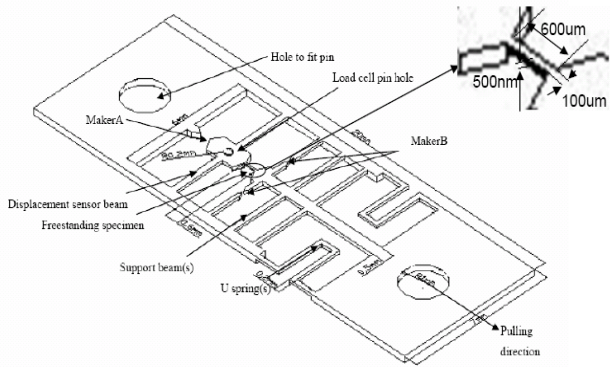

Fig. 2. The schematic of the test chip.

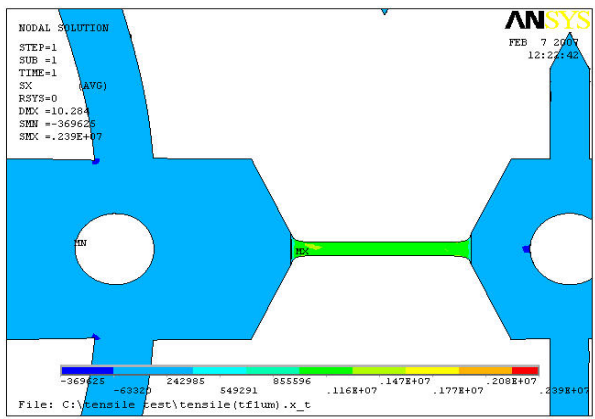

Fig.3. The stress uniformly distributed on the bridge section of thin films while exerting the tensile loading

**2.1 Sample design and fabrication**

The sample fabrication method involves three steps of lithography and two steps of electroplating copper to hold a dog bone freestanding thin film. Using standard wet etching or lift off techniques, a series of microtensile specimens were patterned in metal thin films, holes, and seed layer for spring and frame structure on the underlying silicon oxide coated silicon substrate. Two steps of electroplating processing to distinct spring and frame portion of the test chip. Finally, chemical etched away the silicon oxide to separated electroplated specimen and silicon substrate.

The details of the fabrication procedure are outlined here:
1. RCA clean Si wafer on a double-side-polished 4-inch (diameter) silicon (Si) wafer.
2. Steam furnace growth thin Silicon dioxide layer on double-sides as etching barrier and also scarified layer for the sample.
3. Using standard photolithography techniques on the front side pattern including coating positive photoresist, etch oxide on backside by buffered HF and stripping positive photoresist.
4. Pending one the tested film materials, using sputter, thermal evaporate, CVD or PVD to deposit prefer tested thin film materials on one side of wafer with thickness ranged from 500 ~ 800 nm.
5. Using standard photolithography technique to pattern the test chip on the front side to pattern microbeam structure on the open window, and etch mask on the backside of the wafer.
6. Using specific etch technique either RIE or Chemical enchant to etch thin film materials on one side of wafer to form the pattern of free standing beam spring sample.
7. Pattern etch-mask on the backside side, aligned to cover exactly the gauge length portion of the specimen.
8. Etch 100 microns of silicon substrate by Deep Reactive Ion (ICP).
9. Etching (DRIE or ICP) process of secondary procedure on silicon substrate to form spring and beam section of samples.
10. Remove the photoresist on the backside by O2 plasma.
11. Through the wafer etch from backside to release gauge length, Plasma remove photoresist mask and produce the structural beams.
12. Remove the remaining layer of photoresist and etch away scarified oxide layer by vapor phase buffered HF.

Total of 27 test samples can be obtained from each wafer. A typical fabricated specimen ready for testing is shown in Fig. 4. The freestanding thin film is shown in Figure 5.

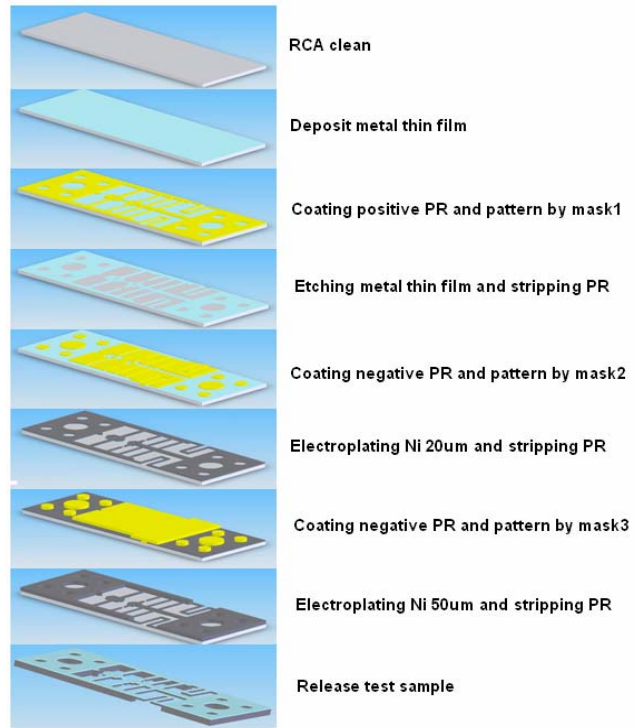

Fig.4. The process schematic of thin film specimen.

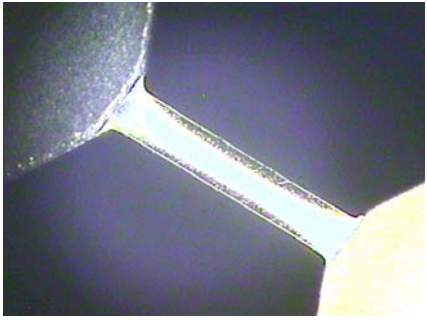

Fig.5. The freestanding thin film.

This left a gage section of free-standing test specimen 600 µm long with a width of 100 µm and a thickness of 200,400, 600 and 800nm. The remaining thin film materials outside the gage section maintained good adhesion to the spring and frame portion of the chip that would subsequently be gripped for tensile testing.

**2.2 Testing system**
The system is custom design equipped with load cells and controlled by PC. This assembly consists of a micromechanical testing system with height-adjustable grip pins, built-in piezoactuator with position sensor, load cell and temperature sensor. The control electronics include a closed-loop piezoelectric controller, amplifier and waveform generator. Monitored signals are conditioned and then fed into an A/D board which is located in a PC. Data acquisition is performed with LabVIEW software. The system is covered in a well insulated, temperature controlled box and is supported on a vibration isolation table. It consists of two aluminum blocks rigidly bolted to the vibration isolation table. The left block mounts a displacement controlled piezoelectric actuator. The right block mounts with height adjustable pin via an x–y–z stage to assure well alignment of the sample and also supports the capacitor load cell.
Simultaneously, external microscope with CCD camera is equipped to acquire in situ image for study during testing. A schematic figure of the micromechanical testing system is shown in Figure 6.
Freestanding thin films were then loaded by performing monotonic loading/unloading up to 2 µm/ (10-2s) and tension-tension fatigue experiments at 5Hz up to ~$10^5$ cycles. Loading was applied using a piezoelectric actuator with 0.1 µm resolution connected through pin hole into the test chip specimen. Loads were measured by connected specimens through a pin connected rod using its bottom pivot to transfer loads and displacement from sample load sensing beam into a capacitor load cell with a resolution of less than 0.1 mN.

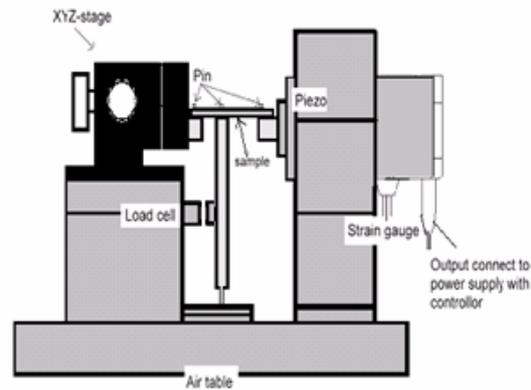

Fig. 6. The schematic figure of testing system.

**3. MEASUREMENTS OF SPECIMEN EXTENSION AND STRAIN**

Measurements of extension and strain of the micrometer sized specimens over the gauge length are difficult to make by means of small scribe lines or marks, or by a clip-on extensometer. White laser interferometry has been used to directly measure the specimen's extension during a test [8], and the digital image-correlation technique to calculate the extension and the strain is another method [9,10]. However, each of the above mention method is expensive. Here, with the usage of the capacitance sensor measure the displacement can be found through the differences of marker A and marker B (see Fig.2). The resolution of the capacitance sensor can reach as accurate as 0.01nm and the elongation of thin films can shown as

$$\delta_f = \Delta y - \Delta x \quad (1)$$

where ∆y and ∆x is displacement of markB and markA respectively.
The force transfer form thin films were calculated by follow equation.

$$F = k_1 \cdot \Delta x \quad (2)$$

where $k_1$ is the spring constant of the force sensor beam. the Young's modulus will calculated by eq.(3)

$$E = \frac{F}{A\delta} \quad (3)$$

where F is the force on the thin film, A is the cross section area of thin films and $\delta$ is the elongation of thin films.

**4. RESULTS**

**4.1 Monotonic testing**
*4.1.1 Thermal evaporate gold*

Fig. 7 shows an example stress-strain plot for thermal evaporate gold tested at room temperature (~25℃), using the procedures described in the testing standards [3,4] we can determine the 0.2% offset yielding strength, the ultimate tensile strength (UTS), the elongation, and the Young's modulus. For the thermal evaporate gold tested at room temperature and average strain rate of $1.67*10^{-4}s^{-1}$, we obtained an average 0.2% offset yield strength of 270MPa, with values ranging from 254 to 283 and an estimated standard deviation of 13.4 MPa.

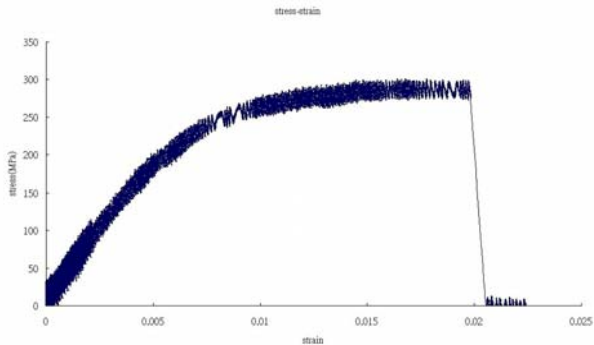

Fig.7. An example stress-strain plot for thermal evaporate gold tested at room temperature

*4.1.2 Sputtering copper*

The validation experiments were then performed on monotonic tensile loading/unloading the specimen corresponding to its strain behavior through three to four random cycles. The results can be observed from Figure 8. The Copper microtensile beam exhibits plasticity prior to failure. The yield stress was determined to be approximately 474.2 MPa and UTS is 575 MPa using an experimentally measured modulus value of 103 GPa in loading and 107 GPa in unloading. The specimen was continuously monitored in the microscope and no sign of curling; cracking or creep was observed up to this point.

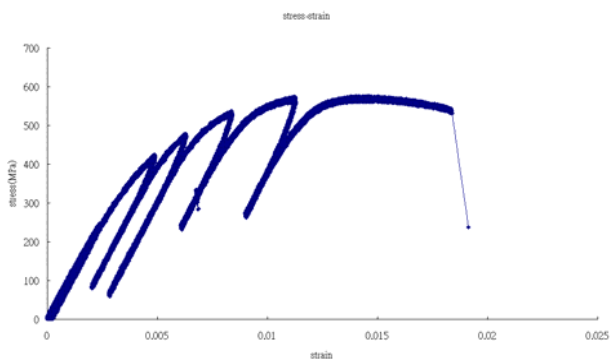

Fig.8. Tensile loading/unloading the specimen corresponding to its strain behavior through three to four random cycles.

**4.2 Fatigue testing**

Tension-tension type fatigue tests were utilized in this study. The experimental parameters were determined based on results from monotonic tensile testing. Two parameters, displacement amplitude (A) and mean displacement (d) were varied in a systematic manner.

Figure 9 and 10 shows the triangular waveform produced by the piezoelectric actuator were varied in the experiments where constant displacement amplitude, Ao was maintained while the mean displacement was varied from do to $d_1$. One experimental constraint to note is that in order to maintain either tension-tension or tension-zero experiments, the displacement amplitude may not be more than two times the mean displacement.

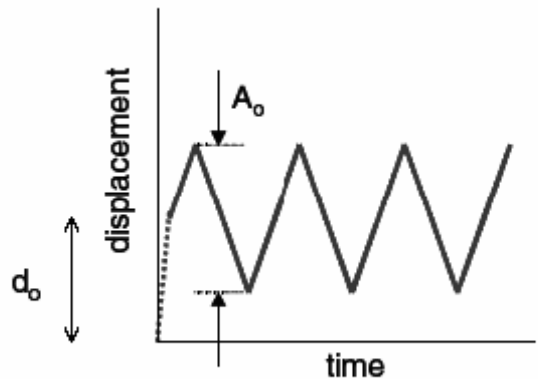

Fig. 9. The schematic of testing method I.

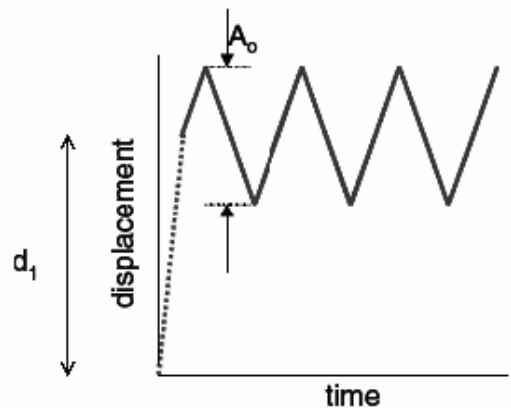

Fig. 10. The schematic of testing method II.

Five fatigue tests set of 300nm copper thin films were performed from mean stress of 100 MPa (0.9 μm) to 300 MPa (2.7 μm) with every 50 MPa (0.45 μm) extensions. Samples were run at a mean displacement corresponding to each load, including displacement amplitude corresponding to +- 20 MPa (0.18 μm) experiment, the maximum amplitude for this particular mean displacement (A/2≤d). In bulk copper materials, under constant stress amplitude, $1 \times 10^2 \sim 10^3$ cycles without failure is often defined as a fatigue limit. However, neither of the samples tested here failed within $3 \times 10^3$ cycles. Although testing conditions are different, we currently consider the 300nm samples to be past the fatigue limit. Table 1 shows the compare between the results tested here and previous fatigue results of copper

|  | Thickness (um) | △ε pl (%) | Yield strength (MPa) | Stress level (MPa) | N of cycles |
|---|---|---|---|---|---|
| D.T. Read | 1.1 |  | 330 | 300 | 1E1~1E3 |
| G.P.Zhang et.al. | 0.4 | 0.86 |  |  | 1E4 |
| R.SCHWAIGER et.al. | 0.4 | 0.83 | 325 |  | 8.8E2 |
| **Test result** | **0.3** | **0.317** | **410** | **300** | **3.3E3** |

thin film.

Table 1 The compare between the results of Cu thin film.

When the experimental mean displacement was increased to 2.7 μm, sample failure was observed at both of the displacement amplitudes which did not fail in the 0.9 μm tests. A summary of the results may be seen in Figure11.

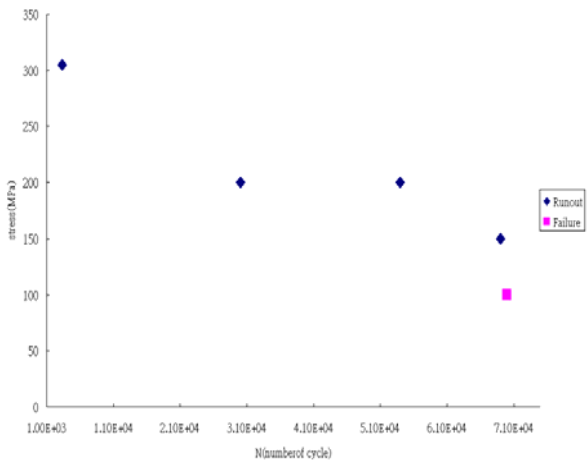

Fig. 11. The S-N curve of 300nm copper thin films.

The fatigue lifetime is dependent on both displacement amplitude and mean displacement. Although it is too early to make quantifiable statements regarding the limited results, a S-N curve was plotted (Figure 11) to fit the data as a starting point for analysis. Further experimental work will include more examination of the effects of displacement amplitude and mean displacement on fatigue life, as well as the effect of film thickness and microstructure on the fatigue properties of thin films.

## 5. CONCLUSION

An experimental sample design and apparatus for uniaxially tensile testing microtensile of thin films under monotonic loading/unloading and tension-tension fatigue conditions has been designed and fabricated.

For the monotonic loading/unloading and stress relaxation experiments, we found the modulus of gold, copper and tantalum nitride thin films with thickness of 200~800nm at ambient temperature. The elastic modulus of thin films obtained here is in close agreement to the value of others. The microtensile beams show plasticity as well as a relaxation dependence on strain rate and stress level. In addition, the yield strength was found to be 1~2 times than the bulk value for copper and gold, which reflects yielding strengthening at this size scale.

For the displacement controlled tension-tension fatigue experiments have also been performed. For fatigue experiments, a trend of decreasing cycles to failure with increasing displacement amplitude and increasing mean displacement has been noted; they also provided proof of the testing approach as well as useful information for the design and development of new MEMS devices.

## ACKNOWLEDGMENTS

This work was supported by Taiwan National Science Council; grant number NSC94-2218-E-005-019.

## REFERENCES

1. Miller, S.L., M.S. Rodgers, G. LaVigne, J.J.Sniegowski, P. Clews, D.M. Tanner, and K.A. Peterson, in Proc. 36th IEEE Int. Reliability Physics Symp. (NJ, 1998)
2. Haque, M.A. and Saif, M.T.A., "In Situ Tensile Testing of nano-scale Specimens in SEM and TEM," EXPERIMENTAL MECHANICS, 42(1), 123-128 (2001)
3. Ming-Tzer Lin, Chi-Jia Tong & Chung-Hsun Chiang "Design and Development of Sub-micron scale Specimens with Electroplated Structures for the Microtensile Testing of Thin Films" MICROSYSTEM TECHNOLOGIES (Accept JAN 2007)
4. Haque, M.A. and Saif, M.T.A., "Application of MEMS force sensors for in situ mechanical characterization of nano-scale thin films in SEM and TEM", Sensors and Actuators A 97-98, 239-245(2002).
5. Ming-Tzer Lin et al. (2006) Design an electroplated framefreestanding specimen for microtensile testing of submicron thin TaN and Cu Film, in materials, technology and reliability of low-k dielectrics and copper interconnects. Mater Res Soc Symp Proc 914, Warrendale

6. D.T. Read "Tension-tension fatigue of copper thin films". Int. J. Fatigue Vol. 20, No. 3, pp. 203-209. (1998)
7. Nicholas Barbosa III, Paul El-Deiry, Richard P. Vinci "Monotonic Testing and Tension-Tension Fatigue Testing of Free-standing Al Microtensile Beams" Mat. Res. Soc. Symp. Proc. Vol. 795 © (2004)
8. D.T.Read ,"Tension-tension fatigue of copper thin films" Int.J.Fatigue Vol. 20.No.3,pp.203-209.1998.
9. R.SCHWAIGER *et.al* "Cyclic deformation of polycrystalline Cu films" PHILOSOPHICAL MAGAZINE,VOL.83,No.6,693-710,2003
10. G.P.Zhang *et.al* "Length-scale-controlled fatigue mechanisms in thin copper films " Acta Materialia 54(2006)3127-3139